\documentclass[12pt,a4paper]{article}
\usepackage[english]{babel}
\usepackage[utf8]{inputenc}
\usepackage[dvips]{graphics}
\usepackage{longtable}
\usepackage{lscape}
\catcode`\{=1
\catcode`\}=2
\def\rfig#1#2#3{figure~#1\begin{figure}[htb]
\includegraphics{#2} \hfil {\centerline {\it figure #1#3}}
\end{figure}}

\def\lref#1{{\tt [#1]}}

\title{NOP -- A Simple Experimental Processor for Parallel Deployment}
\author{Oskar Schirmer}
\date{Göttingen, 2016-12-20}
\begin{document}

\maketitle
\vfill

\section*{Abstract}

The design of a parallel computing system using several thousands
or even up to a million processors asks for processing units that
are simple and thus small in space, to make as many
processing units as possible fit on a single die.

The design presented herewith is far from being optimised,
it is not meant to compete with industry performance devices.
Its main purpose is to allow for a prototypical implementation
of a dynamic software system as a proof of concept.

\vfill
\newpage

\tableofcontents
\newpage

\section{Overview}

The {\it Null Operand Parallel} processor is designed to
build a parallel computing system out of large numbers of
such instances.
Its main purpose is to allow for a prototypical implementation
of a dynamic software system as a proof of concept.

While such processors have been designed, existing approaches
either allow for substantial simplification (\lref{2009dm}),
or do lack flexible communicating means
or sufficient resources (\lref{2011ga})
needed when it comes to implementing a dynamic software system,
i.e. an operating system with user interaction
and dynamic process creation.

The {\it Null Operand Parallel} processor is composed of a
number of processing units, each with its own local fast memory,
and a communication switch to allow exchanging messages
between the processing units, and between processors that are
connected via an external link
(see \rfig{1}{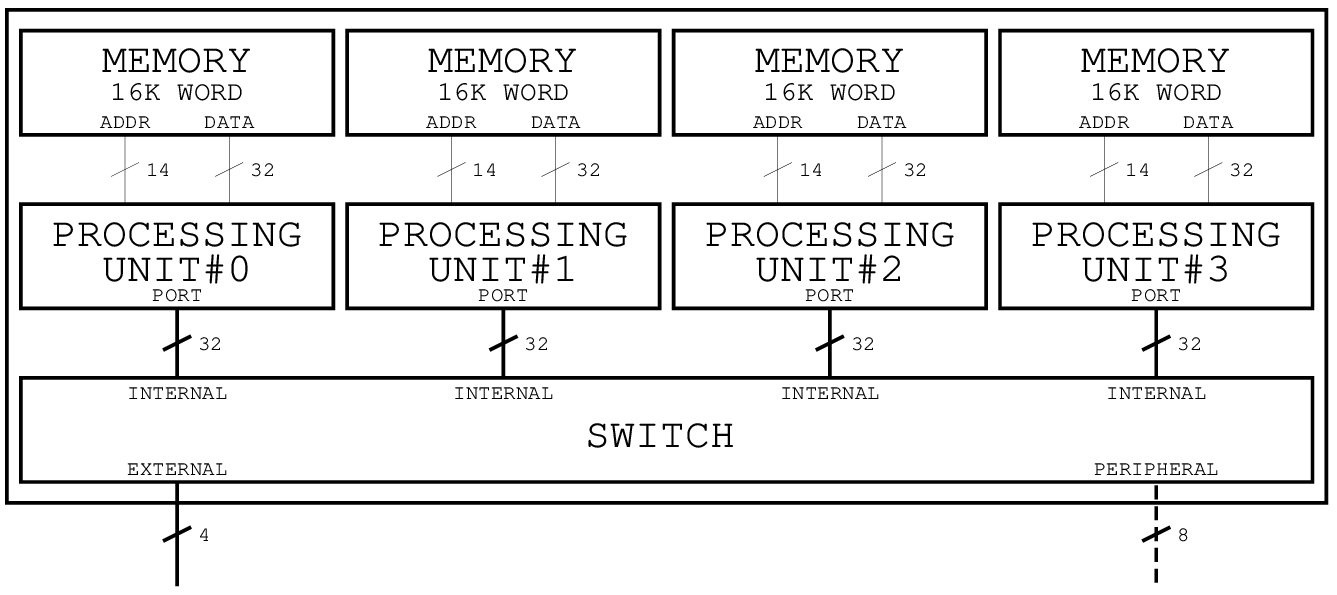}{: NOP block diagram}).

Each processing unit provides multiple register sets to
implement a number of independant hardware threads.
Those threads that are active are scheduled in a simple
round robin manner.
Threads may block -- waiting for data availability --
and stop -- either controlled or upon fault.

All addressing for a single thread is
done relative to base pointers, one for code and constant
data, and one for variable data. This way all operations
are base register relative, and so any thread is fully
relocatable even at runtime.

The single thread is designed to perform sequential
instructions one at a time. Instructions are encoded in
eight bit each, with no operands encoded. Instead,
operations are performed on a single thread local
stack\footnote{which is a well known concept,
see e.g. \lref{1963bc}, see also \lref{2007el}}.

All addressing is done wordwise, i.e. there are no
byte addressable items at all, and consequently, there
are no alignment issues. The only register to address
subunits of words is the instruction pointer, as there
are always four instructions in a word.

For each thread, the processing unit implements a number of
data channel ports that are directly connected
to the communication switch.
The communication switch provides external links,
so it is possible to connect processors and thus build
a large computation network\footnote{Note,
that neither the switch concept is new -- see e.g. \lref{2009dm}
-- nor is the external link concept -- see e.g. \lref{2008es}}.

Transmission of data to a remote port on a different
processor may require using the switches of one or more
intermediate processors, whenever there is no direct
connection between the originating and the target processor.
To handle arbitrary -- and especially transitional -- channels,
the communication switch implements a generic,
table driven routing algorithm.

For peripheral data transmission, the communication switch
provides a set of bidirectional interfaces to transfer
data to and from peripheral units without no specific
transmission protocol.

Numbers for the current test implementation:

\begin{longtable}{|l|l|}
\hline
processing units per processor & 4 \\ \hline
threads per processing units & 8 \\ \hline
word size in bits & 32 \\ \hline
words per local memory & 16384 \\ \hline
channel ports per thread & 32 \\ \hline
external link channels & 4 \\ \hline
peripheral transmission interfaces & 8 \\ \hline
\end{longtable}

The {\it Null Operand Parallel} Processor does not provide:

\begin{description}
\item -- shared memory
\item -- cache memory
\item -- interrupts
\item -- program flow exceptions
\item -- virtual memory addressing
\item -- privileged modes
\item -- dedicated specialised peripheral units
\end{description}

\newpage

\section{Registers and Execution Model}

Execution per thread is based on six registers,
all 14 bit wide, with the exception of the instruction pointer:

\begin{longtable}{|p{.5in}|p{1.3in}|p{3.1in}|}
\hline
ip & instruction pointer &
16 bit wide, the lowest 2 bits referencing the opcode within a word,
least significant opcode first
\\ \hline
sp & stack pointer &
growing downwards
\\ \hline
ld$_n$ & data limits &
lower and upper bound for memory data access
\\ \hline
lc$_n$ & code limits &
lower and upper bound for memory code and constants access
\\ \hline
\end{longtable}

Another two registers are derived from the limits:

\begin{longtable}{|p{.5in}|p{1.3in}|p{3.1in}|}
\hline
cp & constants pointer &
equals lc$_0 + 64$
\\ \hline
dp & data pointer &
equals ld$_0 + 64$
\\ \hline
\end{longtable}

Whenever the instruction pointer is out of range of
the code limits, or the stack pointer is out of range
of the data limits, the thread is stopped as faulty
(see \rfig{2}{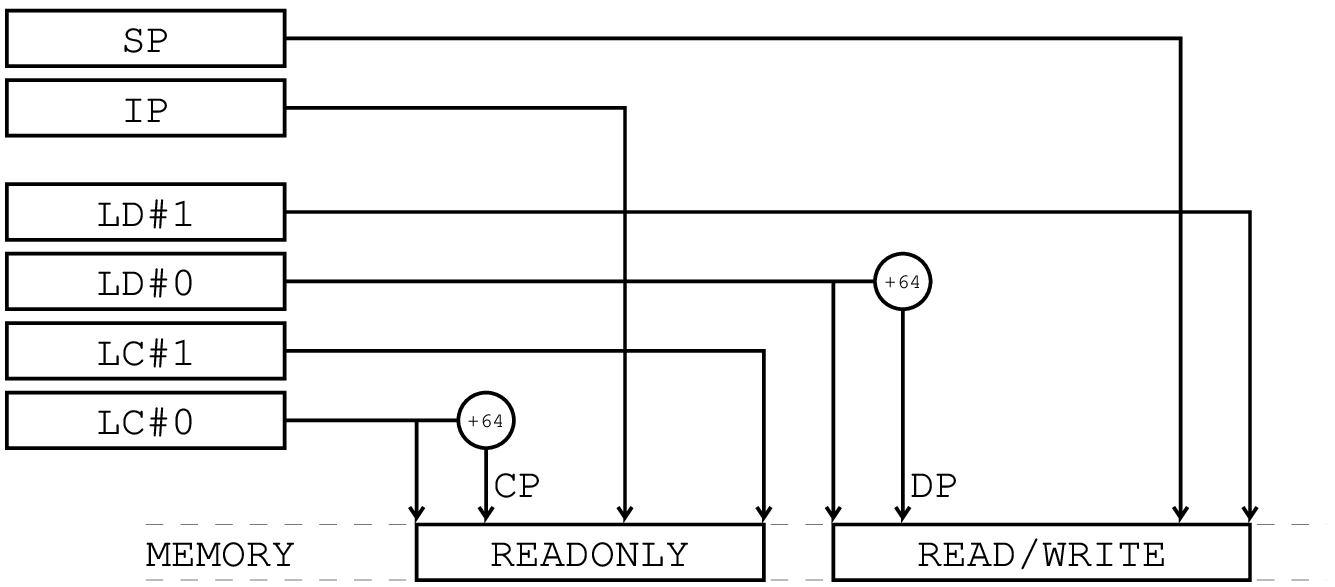}{: NOP register model}).

At start time of a thread, its stack pointer is
initialised to ld$_1$.

When a thread is stopped -- be it faulty or not --
an {\it exception message} is sent to a destination port,
stored into the {\it exc} register at thread start time.

\newpage

\section{Communication Switch}

To send a message through a channel,
first the destination channel port number has to be assigned
to the local channel port. Any subsequent message send through
this channel is send to that destination, until a new destination
is assigned. A channel global port number is 32 bit wide:

\begin{longtable}{|p{2.6in}|p{.7in}|p{.7in}|p{.7in}|}
\hline
processor id resp. routing command & unit\# & thread\# & port\# \\ \hline
bit 31 .. 10 & bit 9 .. 8 & bit 7 .. 5 & bit 4 .. 0 \\ \hline
\end{longtable}

For special values of the upper 22 bits of the global port number,
special local routing is performed:

\begin{longtable}{|p{2.2in}|p{2.8in}|}
\hline
routing command value & routing action \\ \hline
0 & connect to local unit \\ \hline
1 & connect to peripheral line \\ \hline
2 & connect to router configuration block \\ \hline
4 .. 7 & connect to external link 0 .. 3 \\ \hline
processor id $\geq$ 8 & connect according to routing table \\ \hline
\end{longtable}

A channel transmission path in use is blocked from its local
to the destination end as long as the message is sent.
To end a message, the sender
must send a terminatory {\it END} token. When sending a message is
paused and shall be continued later, the sender is free to send
in between a {\it PAUSE} token, which -- in contrast to the {\it END} token --
is not delivered to the receiving destination. By sending a
{\it PAUSE} token, the channel transmission path is freed until sending
the message is continued.

\newpage

\section{Instruction Opcodes}

All instructions are encoded as eight bit opcodes,
and do not encode operands explicitely but work on stack data.

Boolean values are words, {\tt 0} is false, all other values are true.
Instructions, that generate a boolean value, will always push
{\tt -1} for true.

In the following detailed description of the single
instructions, {\tt m} denotes the local memory.
Other single character variables denote temporary values.
{\tt --sp} is the stack pointer predecremented before use,
{\tt sp++} is the stack pointer postincremented after use.

The {\tt port} function calculates a global port number from the
processors id, the processing units number, the given thread number,
and the local channel port number:

\begin{tt}
\noindent
port(t, p)$_{31..10}$ $\leftarrow$ id$_{processor}$\\
port(t, p)$_{9..8}$ $\leftarrow$ number$_{unit}$\\
port(t, p)$_{7..5}$ $\leftarrow$ t\\
port(t, p)$_{4..0}$ $\leftarrow$ p
\end{tt}

\subsection{Immediate Constant}

\begin{longtable}{|p{5.2in}|}
\hline
Opcodes: {\tt 0x00}..{\tt 0x7F}, {\tt 0xC0}..{\tt 0xFF}
\\ \hline
\end{longtable}

All opcodes, except those in the range from 0x80 to 0xBF,
do load their own constant value -- sign extended -- onto the stack:

\begin{tt}
\noindent
a$_{31..8}$ $\leftarrow$ opcode$_7$\\
a$_{7..0}$ $\leftarrow$ opcode\\
m(--sp) $\leftarrow$ a\\
ip $\leftarrow$ ip + 1
\end{tt}

\subsection{No Operation}

\begin{longtable}{|p{2.5in}|p{2.5in}|}
\hline
Name: NOP &
Opcode: {\tt 0x80} \\ \hline
\end{longtable}

No operation:

\begin{tt}
\noindent
ip $\leftarrow$ ip + 1
\end{tt}

\subsection{Add}

\begin{longtable}{|p{2.5in}|p{2.5in}|}
\hline
Name: ADD &
Opcode: {\tt 0x81} \\ \hline
\end{longtable}

Pop two words, push the sum:

\begin{tt}
\noindent
a $\leftarrow$ m(sp++)\\
b $\leftarrow$ m(sp++)\\
m(--sp) $\leftarrow$ b + a\\
ip $\leftarrow$ ip + 1
\end{tt}

\subsection{Subtract}

\begin{longtable}{|p{2.5in}|p{2.5in}|}
\hline
Name: SUB &
Opcode: {\tt 0x82} \\ \hline
\end{longtable}

Pop a subtrahend, pop a minuend, subtract the subtrahend from the minuend,
push difference:

\begin{tt}
\noindent
a $\leftarrow$ m(sp++)\\
b $\leftarrow$ m(sp++)\\
m(--sp) $\leftarrow$ b - a\\
ip $\leftarrow$ ip + 1
\end{tt}

\subsection{Multiply}

\begin{longtable}{|p{2.5in}|p{2.5in}|}
\hline
Name: MUL &
Opcode: {\tt 0x83} \\ \hline
\end{longtable}

Pop two words, push the product:

\begin{tt}
\noindent
a $\leftarrow$ m(sp++)\\
b $\leftarrow$ m(sp++)\\
m(--sp) $\leftarrow$ b * a\\
ip $\leftarrow$ ip + 1
\end{tt}

\subsection{Unsigned Divide}

\begin{longtable}{|p{2.5in}|p{2.5in}|}
\hline
Name: UDIV &
Opcode: {\tt 0x84} \\ \hline
\end{longtable}

Pop a divisor, pop a divident, unsigned divide divident by divisor,
push quotient and remainder.
On division by zero stop thread:

\begin{tt}
\noindent
a $\leftarrow$ m(sp++)\\
if a = 0 then fault\\
b $\leftarrow$ m(sp++)\\
q $\leftarrow$ b / a\\
m(--sp) $\leftarrow$ q\\
m(--sp) $\leftarrow$ b - q * a\\
ip $\leftarrow$ ip + 1
\end{tt}

\subsection{Signed Divide}

\begin{longtable}{|p{2.5in}|p{2.5in}|}
\hline
Name: SDIV &
Opcode: {\tt 0x85} \\ \hline
\end{longtable}

Pop a divisor, pop a divident, signed divide divident by divisor
according to the Euclidean division (see e.g. \lref{2001dl}),
push quotient and remainder.
On division by zero stop thread:

\begin{tt}
\noindent
a $\leftarrow_{two's~complement}$ m(sp++)\\
if a = 0 then fault\\
b $\leftarrow_{two's~complement}$ m(sp++)\\
q $\leftarrow$ b / a\\
m(--sp) $\leftarrow$ q\\
m(--sp) $\leftarrow$ b - q * a\\
ip $\leftarrow$ ip + 1
\end{tt}

\subsection{Bitwise And}

\begin{longtable}{|p{2.5in}|p{2.5in}|}
\hline
Name: AND &
Opcode: {\tt 0x86} \\ \hline
\end{longtable}

Pop two words, push the conjunction:

\begin{tt}
\noindent
a $\leftarrow$ m(sp++)\\
b $\leftarrow$ m(sp++)\\
m(--sp) $\leftarrow$ b and a\\
ip $\leftarrow$ ip + 1
\end{tt}

\subsection{Bitwise Or}

\begin{longtable}{|p{2.5in}|p{2.5in}|}
\hline
Name: OR &
Opcode: {\tt 0x87} \\ \hline
\end{longtable}

Pop two words, push the disjunction:

\begin{tt}
\noindent
a $\leftarrow$ m(sp++)\\
b $\leftarrow$ m(sp++)\\
m(--sp) $\leftarrow$ b or a\\
ip $\leftarrow$ ip + 1
\end{tt}

\subsection{Bitwise Exclusive Or}

\begin{longtable}{|p{2.5in}|p{2.5in}|}
\hline
Name: XOR &
Opcode: {\tt 0x88} \\ \hline
\end{longtable}

pop two words, push the exclusion:

\begin{tt}
\noindent
a $\leftarrow$ m(sp++)\\
b $\leftarrow$ m(sp++)\\
m(--sp) $\leftarrow$ b xor a\\
ip $\leftarrow$ ip + 1
\end{tt}

\subsection{Pop}

\begin{longtable}{|p{2.5in}|p{2.5in}|}
\hline
Name: POP &
Opcode: {\tt 0x89} \\ \hline
\end{longtable}

Pop one word and discard it:

\begin{tt}
\noindent
sp $\leftarrow$ sp + 1\\
ip $\leftarrow$ ip + 1
\end{tt}

\subsection{Duplicate}

\begin{longtable}{|p{2.5in}|p{2.5in}|}
\hline
Name: DUP &
Opcode: {\tt 0x8A} \\ \hline
\end{longtable}

Pop one word and push it twice:

\begin{tt}
\noindent
a $\leftarrow$ m(sp)\\
m(--sp) $\leftarrow$ a\\
ip $\leftarrow$ ip + 1
\end{tt}

\subsection{Exchange}

\begin{longtable}{|p{2.5in}|p{2.5in}|}
\hline
Name: EXCH &
Opcode: {\tt 0x8B} \\ \hline
\end{longtable}

Pop one word, pop another, then push the one, push the other:

\begin{tt}
\noindent
a $\leftarrow$ m(sp++)\\
b $\leftarrow$ m(sp++)\\
m(--sp) $\leftarrow$ a\\
m(--sp) $\leftarrow$ b\\
ip $\leftarrow$ ip + 1
\end{tt}

\subsection{Load From Stack}

\begin{longtable}{|p{2.5in}|p{2.5in}|}
\hline
Name: LDX &
Opcode: {\tt 0x8C} \\ \hline
\end{longtable}

Pop an index, add it to the stack pointer, push word it points to:

\begin{tt}
\noindent
a $\leftarrow$ m(sp++)\\
b $\leftarrow$ m(sp + a)\\
m(--sp) $\leftarrow$ b\\
ip $\leftarrow$ ip + 1
\end{tt}

\subsection{Swap Bit Fields}

\begin{longtable}{|p{2.5in}|p{2.5in}|}
\hline
Name: SWAP &
Opcode: {\tt 0x8D} \\ \hline
\end{longtable}

pop a 5 bit mask, pop a word, for each set bit i in the mask reverse all
32 bits in a word, exchanging each bit with the bit at offset $2^i$.
Use mask 24 for endianess swap, use mask 31 for full bitwise reversal:

\begin{tt}
\noindent
a $\leftarrow$ m(sp++)$_{4..0}$\\
b $\leftarrow$ m(sp++)\\
for i in 0..4 do\\
\hspace*{2ex} if a$_i$ = 1 then\\
\hspace*{4ex} for k in 0..31 do\\
\hspace*{6ex} b'$_k$ $\leftarrow$ b$_{(k~xor~(2^i))}$\\
\hspace*{4ex} b $\leftarrow$ b'\\
ip $\leftarrow$ ip + 1
\end{tt}

\subsection{Load, Decrement, Push, Store}

\begin{longtable}{|p{2.5in}|p{2.5in}|}
\hline
Name: DECLD &
Opcode: {\tt 0x8E} \\ \hline
\end{longtable}

Pop an index, add it to the data pointer, load the word it points to,
decrement it, push it, store it back:

\begin{tt}
\noindent
a $\leftarrow$ m(sp++)\\
b $\leftarrow$ m(a + dp)\\
b $\leftarrow$ b - 1\\
m(--sp) $\leftarrow$ b\\
m(a + dp) $\leftarrow$ b\\
ip $\leftarrow$ ip + 1
\end{tt}

\subsection{Logarithm}

\begin{longtable}{|p{2.5in}|p{2.5in}|}
\hline
Name: LOG2 &
Opcode: {\tt 0x8F} \\ \hline
\end{longtable}

Determine position of highest bit set, -1 for zero word:

\begin{tt}
\noindent
a $\leftarrow$ m(sp++)\\
if a = 0 then\\
\hspace*{2ex} b $\leftarrow$ -1\\
else\\
\hspace*{2ex} b $\leftarrow$ $i$ $\vert$ a$_i$ = 1 $\wedge$ $\forall j>i:$ a$_j$ = 0\\
ip $\leftarrow$ ip + 1
\end{tt}

\subsection{Shift And Rotate Left}

\begin{longtable}{|p{2.5in}|p{2.5in}|}
\hline
Name: LEFT &
Opcode: {\tt 0x90} \\ \hline
\end{longtable}

Pop a bit count, pop a rotator word and a shifter word,
shift the shifter by count bits to left, shifting in the bits from the
rotator, which is rotated to the left by count bits, push the shifter result:

\begin{tt}
\noindent
a $\leftarrow$ m(sp++)\\
b $\leftarrow$ m(sp++)\\
c $\leftarrow$ m(sp++)\\
a' $\leftarrow$ a$_{4..0}$\\
if a < 32 then\\
\hspace*{2ex} c'$_{31..a'}$ $\leftarrow$ c$_{(31-a')..0}$\\
else\\
\hspace*{2ex} c'$_{31..a'}$ $\leftarrow$ b$_{(31-a')..0}$\\
c'$_{(a'-1)..0}$ $\leftarrow$ b$_{31..(32-a')}$\\
m(--sp) $\leftarrow$ c'\\
ip $\leftarrow$ ip + 1
\end{tt}

\subsection{Shift And Rotate Right}

\begin{longtable}{|p{2.5in}|p{2.5in}|}
\hline
Name: RIGHT &
Opcode: {\tt 0x91} \\ \hline
\end{longtable}

Pop a bit count n, pop a rotator word and a shifter word,
shift the shifter by n bits to right, shifting in the bits from the
rotator, which is rotated to the right by n bits, push the shifter result:

\begin{tt}
\noindent
a $\leftarrow$ m(sp++)\\
b $\leftarrow$ m(sp++)\\
c $\leftarrow$ m(sp++)\\
a' $\leftarrow$ a$_{4..0}$\\
if a < 32 then\\
\hspace*{2ex} c'$_{(31-a')..0}$ $\leftarrow$ c$_{31..a'}$\\
else\\
\hspace*{2ex} c'$_{(31-a')..0}$ $\leftarrow$ b$_{31..a'}$\\
c'$_{31..(32-a')}$ $\leftarrow$ b$_{(a'-1)..0}$\\
m(--sp) $\leftarrow$ c'\\
ip $\leftarrow$ ip + 1
\end{tt}

\subsection{Sign Extend}

\begin{longtable}{|p{2.5in}|p{2.5in}|}
\hline
Name: SIGN &
Opcode: {\tt 0x92} \\ \hline
\end{longtable}

Pop a word, push true if it is negative, false otherwise:

\begin{tt}
\noindent
a $\leftarrow$ m(sp++)\\
a$_{31..1}$ $\leftarrow$ a$_0$\\
m(--sp) $\leftarrow$ a\\
ip $\leftarrow$ ip + 1
\end{tt}

\subsection{Zero}

\begin{longtable}{|p{2.5in}|p{2.5in}|}
\hline
Name: ZERO &
Opcode: {\tt 0x93} \\ \hline
\end{longtable}

pop a word, push true if it is zero, false otherwise:

\begin{tt}
\noindent
a $\leftarrow$ m(sp++)\\
if a = 0 then\\
\hspace*{2ex} b $\leftarrow$ -1\\
else\\
\hspace*{2ex} b $\leftarrow$ 0\\
m(--sp) $\leftarrow$ b\\
ip $\leftarrow$ ip + 1
\end{tt}

\subsection{Unconditional Jump}

\begin{longtable}{|p{2.5in}|p{2.5in}|}
\hline
Name: UJP &
Opcode: {\tt 0x94} \\ \hline
\end{longtable}

Pop an offset, add it to the current instruction pointer:

\begin{tt}
\noindent
a $\leftarrow$ m(sp++)\\
ip $\leftarrow$ ip + a
\end{tt}

\subsection{False Jump}

\begin{longtable}{|p{2.5in}|p{2.5in}|}
\hline
Name: FJP &
Opcode: {\tt 0x95} \\ \hline
\end{longtable}

Pop an offset and a condition, if condition is false, add the offset
to the current instruction pointer:

\begin{tt}
\noindent
a $\leftarrow$ m(sp++)\\
b $\leftarrow$ m(sp++)\\
if b = 0 then\\
\hspace*{2ex} ip $\leftarrow$ ip + a\\
else\\
\hspace*{2ex} ip $\leftarrow$ ip + 1
\end{tt}

\subsection{Load Constant}

\begin{longtable}{|p{2.5in}|p{2.5in}|}
\hline
Name: LDC &
Opcode: {\tt 0x96} \\ \hline
\end{longtable}

Pop an index, add it to the constant pointer, push word it points to.
However, if that pointer is out of range of the constant limits, then
subtract the constant pool size (lc$_1$ - lc$_0$)
and add the data pointer instead:

\begin{tt}
\noindent
a $\leftarrow$ m(sp++)\\
if (a + cp) $\geq$ lc$_1$ then\\
\hspace*{2ex} b $\leftarrow$ a - $($lc$_1$ - lc$_0)$ + dp\\
else\\
\hspace*{2ex} b $\leftarrow$ a + cp\\
m(--sp) $\leftarrow$ m(b)\\
ip $\leftarrow$ ip + 1
\end{tt}

\subsection{Load Word}

\begin{longtable}{|p{2.5in}|p{2.5in}|}
\hline
Name: LD &
Opcode: {\tt 0x97} \\ \hline
\end{longtable}

Pop an index, add it to the data pointer, push word it points to:

\begin{tt}
\noindent
a $\leftarrow$ m(sp++)\\
m(--sp) $\leftarrow$ m(a + dp)\\
ip $\leftarrow$ ip + 1
\end{tt}

\subsection{Store Word}

\begin{longtable}{|p{2.5in}|p{2.5in}|}
\hline
Name: ST &
Opcode: {\tt 0x98} \\ \hline
\end{longtable}

Pop an index, add it to the data pointer, pop a word and store it:

\begin{tt}
\noindent
a $\leftarrow$ m(sp++)\\
m(a + dp) $\leftarrow$ m(sp++)\\
ip $\leftarrow$ ip + 1
\end{tt}

\subsection{Count Bits}

\begin{longtable}{|p{2.5in}|p{2.5in}|}
\hline
Name: COUNT &
Opcode: {\tt 0x99} \\ \hline
\end{longtable}

Pop a word, count the number of bits set:

\begin{tt}
\noindent
a $\leftarrow$ m(sp++)\\
b $\leftarrow$ $\sum_{i=0..31}($a$_i)$\\
ip $\leftarrow$ ip + 1
\end{tt}

\subsection{Stop}

\begin{longtable}{|p{2.5in}|p{2.5in}|}
\hline
Name: STOP &
Opcode: {\tt 0x9A} \\ \hline
\end{longtable}

Stop the thread:

\begin{tt}
\noindent
stop
\end{tt}

\subsection{Break}

\begin{longtable}{|p{2.5in}|p{2.5in}|}
\hline
Name: BREAK &
Opcode: {\tt 0x9B} \\ \hline
\end{longtable}

Do nothing, but stall for debugging:

\begin{tt}
\noindent
ip $\leftarrow$ ip + 1
\end{tt}

\subsection{Start New Thread}

\begin{longtable}{|p{2.5in}|p{2.5in}|}
\hline
Name: START &
Opcode: {\tt 0x9C} \\ \hline
\end{longtable}

Start a new thread with given limits, ip,
and exception destination port,
set sp to ld$_1$, dp to ld$_0$+64,
and cp to lc$_0$+64, push port id of new threads control port:

\begin{tt}
\noindent
if all threads in use then fault\\
exc $\leftarrow$ m(sp++)\\
ld$_1$' $\leftarrow$ m(sp++)\\
ld$_0$' $\leftarrow$ m(sp++)\\
b $\leftarrow$ m(sp++)\\
lc$_1$' $\leftarrow$ m(sp++)\\
lc$_0$' $\leftarrow$ m(sp++)\\
ip' $\leftarrow$ (b + lc$_0$') * 4\\
sp' $\leftarrow$ ld$_1$'\\
a $\leftarrow$ port(thread', 0)\\
m(--sp) $\leftarrow$ a\\
ip $\leftarrow$ ip + 1
\end{tt}

\subsection{Call Function}

\begin{longtable}{|p{2.5in}|p{2.5in}|}
\hline
Name: CALL &
Opcode: {\tt 0x9D} \\ \hline
\end{longtable}

Pop an offset, add it to the current instruction pointer,
save return address to the stack:

\begin{tt}
\noindent
a $\leftarrow$ m(sp++)\\
m(--sp) $\leftarrow$ ip + 1 - lc$_0$\\
ip $\leftarrow$ ip + a
\end{tt}

\subsection{Return From Function}

\begin{longtable}{|p{2.5in}|p{2.5in}|}
\hline
Name: JUMP &
Opcode: {\tt 0x9E} \\ \hline
\end{longtable}

Restore the instruction pointer from the stack:

\begin{tt}
\noindent
a $\leftarrow$ m(sp++)\\
ip $\leftarrow$ a + lc$_0$\\
\end{tt}

\subsection{Store To Stack}

\begin{longtable}{|p{2.5in}|p{2.5in}|}
\hline
Name: STX &
Opcode: {\tt 0x9F} \\ \hline
\end{longtable}

Pop an index, pop a word, add the index to the stack pointer, store word:

\begin{tt}
\noindent
a $\leftarrow$ m(sp++)\\
b $\leftarrow$ m(sp++)\\
m(sp + a) $\leftarrow$ b\\
ip $\leftarrow$ ip + 1
\end{tt}

\subsection{Load, Push, Increment, Store}

\begin{longtable}{|p{2.5in}|p{2.5in}|}
\hline
Name: LDINC &
Opcode: {\tt 0xA0} \\ \hline
\end{longtable}

Pop an index, add it to the data pointer, load the word it points to,
push it, increment it, store it back:

\begin{tt}
\noindent
a $\leftarrow$ m(sp++)\\
b $\leftarrow$ m(a + dp)\\
m(--sp) $\leftarrow$ b\\
b $\leftarrow$ b + 1\\
m(a + dp) $\leftarrow$ b\\
ip $\leftarrow$ ip + 1
\end{tt}

\subsection{Get Port Destination}

\begin{longtable}{|p{2.5in}|p{2.5in}|}
\hline
Name: GETPORT &
Opcode: {\tt 0xA1} \\ \hline
\end{longtable}

Pop local port id, push local ports destination:

\begin{tt}
\noindent
a $\leftarrow$ m(sp++)\\
b $\leftarrow$ dest$_a$\\
m(--sp) $\leftarrow$ b\\
ip $\leftarrow$ ip + 1
\end{tt}

\subsection{Set Port Destination}

\begin{longtable}{|p{2.5in}|p{2.5in}|}
\hline
Name: SETPORT &
Opcode: {\tt 0xA2} \\ \hline
\end{longtable}

Pop local port id and remote port id, set local ports destination:

\begin{tt}
\noindent
a $\leftarrow$ m(sp++)\\
b $\leftarrow$ m(sp++)\\
dest$_a$ $\leftarrow$ b\\
ip $\leftarrow$ ip + 1
\end{tt}

\subsection{Output Word}

\begin{longtable}{|p{2.5in}|p{2.5in}|}
\hline
Name: OUT &
Opcode: {\tt 0xA3} \\ \hline
\end{longtable}

Pop local port id and a data word, send data into port; may block:

\begin{tt}
\noindent
a $\leftarrow$ m(sp++)\\
b $\leftarrow$ m(sp++)\\
out$_a$ $\leftarrow$ b\\
ip $\leftarrow$ ip + 1
\end{tt}

\subsection{Output End Token}

\begin{longtable}{|p{2.5in}|p{2.5in}|}
\hline
Name: OUTEND &
Opcode: {\tt 0xA4} \\ \hline
\end{longtable}

Pop local port id, send an {\it END} token into port; may block:

\begin{tt}
\noindent
a $\leftarrow$ m(sp++)\\
out$_a$ $\leftarrow$ {\it END}\\
ip $\leftarrow$ ip + 1
\end{tt}

\subsection{Output Pause Token}

\begin{longtable}{|p{2.5in}|p{2.5in}|}
\hline
Name: OUTPAUSE &
Opcode: {\tt 0xA5} \\ \hline
\end{longtable}

Pop local port id, send a {\it PAUSE} token into port; may block:

\begin{tt}
\noindent
a $\leftarrow$ m(sp++)\\
out$_a$ $\leftarrow$ {\it PAUSE}\\
ip $\leftarrow$ ip + 1
\end{tt}

\subsection{Input Word}

\begin{longtable}{|p{2.5in}|p{2.5in}|}
\hline
Name: IN &
Opcode: {\tt 0xA6} \\ \hline
\end{longtable}

Pop local port id, receive a data word from port, push it; may block:

\begin{tt}
\noindent
a $\leftarrow$ m(sp++)\\
b $\leftarrow$ in$_a$\\
if b = {\it END} then fault\\
m(--sp) $\leftarrow$ b\\
ip $\leftarrow$ ip + 1
\end{tt}

\subsection{Check Input Port}

\begin{longtable}{|p{2.5in}|p{2.5in}|}
\hline
Name: INMORE &
Opcode: {\tt 0xA7} \\ \hline
\end{longtable}

Pop local port id, check whether end token was received, discard end token
and push zero if so, push non-zero otherwise; may block:

\begin{tt}
\noindent
a $\leftarrow$ m(sp++)\\
if in$_a$ = {\it END} then\\
\hspace*{2ex} b $\leftarrow$ in$_a$\\
\hspace*{2ex} b $\leftarrow$ 0\\
else\\
\hspace*{2ex} b $\leftarrow$ -1\\
m(--sp) $\leftarrow$ b\\
ip $\leftarrow$ ip + 1
\end{tt}

\subsection{Clear Event List}

\begin{longtable}{|p{2.5in}|p{2.5in}|}
\hline
Name: EVCLEAR &
Opcode: {\tt 0xA8} \\ \hline
\end{longtable}

Clear the event list for current thread:

\begin{tt}
\noindent
ev$_{*,*}$ $\leftarrow$ $\epsilon$\\
ip $\leftarrow$ ip + 1
\end{tt}

\subsection{Set Output Event}

\begin{longtable}{|p{2.5in}|p{2.5in}|}
\hline
Name: EVOUT &
Opcode: {\tt 0xA9} \\ \hline
\end{longtable}

Pop local port id, add event handle for output availability,
pop an offset, add it to the current instruction pointer, use as event vector:

\begin{tt}
\noindent
a $\leftarrow$ m(sp++)\\
b $\leftarrow$ m(sp++)\\
ev$_{a,out}$ $\leftarrow$ b + ip\\
ip $\leftarrow$ ip + 1
\end{tt}

\subsection{Set Input Event}

\begin{longtable}{|p{2.5in}|p{2.5in}|}
\hline
Name: EVIN &
Opcode: {\tt 0xAA} \\ \hline
\end{longtable}

Pop local port id, add event handle for input availability,
pop an offset, add it to the current instruction pointer, use as event vector:

\begin{tt}
\noindent
a $\leftarrow$ m(sp++)\\
b $\leftarrow$ m(sp++)\\
ev$_{a,in}$ $\leftarrow$ b + ip\\
ev$_{a,end}$ $\leftarrow$ b + ip\\
ip $\leftarrow$ ip + 1
\end{tt}

\subsection{Set Input End Event}

\begin{longtable}{|p{2.5in}|p{2.5in}|}
\hline
Name: EVEND &
Opcode: {\tt 0xAB} \\ \hline
\end{longtable}

Pop local port id, add event handle for input end token availability,
pop an offset, add it to the current instruction pointer, use as event vector:

\begin{tt}
\noindent
a $\leftarrow$ m(sp++)\\
b $\leftarrow$ m(sp++)\\
ev$_{a,end}$ $\leftarrow$ b + ip\\
ip $\leftarrow$ ip + 1
\end{tt}

\subsection{Wait}

\begin{longtable}{|p{2.5in}|p{2.5in}|}
\hline
Name: WAIT &
Opcode: {\tt 0xAC} \\ \hline
\end{longtable}

Wait for any of the configured events. as soon as any one occurs, load the
corresponding event vector as new instruction pointer; may block:

\begin{tt}
\noindent
if out$_i$ = $\epsilon$ and ev$_{a,out}$ then\\
\hspace*{2ex} ip $\leftarrow$ ev$_{a,out}$\\
or if in$_i$ = {\it END} and ev$_{a,end}$ then\\
\hspace*{2ex} ip $\leftarrow$ ev$_{a,end}$\\
else if in$_i$ $\neq$ $\epsilon$ and ev$_{a,in}$ then\\
\hspace*{2ex} ip $\leftarrow$ ev$_{a,in}$\\
else\\
\hspace*{2ex} wait
\end{tt}

\subsection{Current Time}

\begin{longtable}{|p{2.5in}|p{2.5in}|}
\hline
Name: NOW &
Opcode: {\tt 0xAD} \\ \hline
\end{longtable}

Push the current time counter:

\begin{tt}
\noindent
m(--sp) $\leftarrow$ {\it time}\\
ip $\leftarrow$ ip + 1
\end{tt}

\subsection{Wait With Timeout}

\begin{longtable}{|p{2.5in}|p{2.5in}|}
\hline
Name: WAITTMO &
Opcode: {\tt 0xAE} \\ \hline
\end{longtable}

Pop a time, wait for any of the configured events. As soon as any one occurs,
load the corresponding event vector as new instruction pointer. When no event
occurs until the specified time is reached, continue instruction execution
without branching; may block:

\begin{tt}
\noindent
a $\leftarrow$ m(sp++)\\
if {\it time} - a $\geq$ 0 then\\
\hspace*{2ex} ip $\leftarrow$ ip + 1\\
or if out$_i$ = $\epsilon$ and ev$_{a,out}$ then\\
\hspace*{2ex} ip $\leftarrow$ ev$_{a,out}$\\
or if in$_i$ = {\it END} and ev$_{a,end}$ then\\
\hspace*{2ex} ip $\leftarrow$ ev$_{a,end}$\\
else if in$_i$ $\neq$ $\epsilon$ and ev$_{a,in}$ then\\
\hspace*{2ex} ip $\leftarrow$ ev$_{a,in}$\\
else\\
\hspace*{2ex} wait
\end{tt}

\subsection{Increment Stack Pointer}

\begin{longtable}{|p{2.5in}|p{2.5in}|}
\hline
Name: POPN &
Opcode: {\tt 0xAF} \\ \hline
\end{longtable}

Pop a summand, add it to the stack pointer:

\begin{tt}
\noindent
a $\leftarrow$ m(sp++)\\
sp $\leftarrow$ sp + a\\
ip $\leftarrow$ ip + 1
\end{tt}

\subsection{Compare Unsigned}

\begin{longtable}{|p{2.5in}|p{2.5in}|}
\hline
Name: ULESS &
Opcode: {\tt 0xB0} \\ \hline
\end{longtable}

Pop a subtrahend, pop a minuend, subtract the subtrahend from the minuend,
push true if negative, false otherwise, all unsigned:

\begin{tt}
\noindent
a $\leftarrow$ m(sp++)\\
b $\leftarrow$ m(sp++)\\
if b < a then\\
\hspace*{2ex} c $\leftarrow$ -1\\
else\\
\hspace*{2ex} c $\leftarrow$ 0\\
m(--sp) $\leftarrow$ c\\
ip $\leftarrow$ ip + 1
\end{tt}

\subsection{Compare Signed}

\begin{longtable}{|p{2.5in}|p{2.5in}|}
\hline
Name: SLESS &
Opcode: {\tt 0xB1} \\ \hline
\end{longtable}

Pop a subtrahend, pop a minuend, subtract the subtrahend from the minuend,
push true if negative, false otherwise, all signed:

\begin{tt}
\noindent
a $\leftarrow_{two's~complement}$ m(sp++)\\
b $\leftarrow_{two's~complement}$ m(sp++)\\
if b < a then\\
\hspace*{2ex} c $\leftarrow$ -1\\
else\\
\hspace*{2ex} c $\leftarrow$ 0\\
m(--sp) $\leftarrow$ c\\
ip $\leftarrow$ ip + 1
\end{tt}

\subsection{Combined Number}

\begin{longtable}{|p{2.5in}|p{2.5in}|}
\hline
Name: COMBINE &
Opcode: {\tt 0xB2} \\ \hline
\end{longtable}

Pop a word, pop another word, multiply the latter by 192, push the sum:

\begin{tt}
\noindent
a $\leftarrow$ m(sp++)\\
b $\leftarrow$ m(sp++)\\
m(--sp) $\leftarrow$ b * 192 + a\\
ip $\leftarrow$ ip + 1
\end{tt}

\subsection{Calculate Global Port Number}

\begin{longtable}{|p{2.5in}|p{2.5in}|}
\hline
Name: PORT &
Opcode: {\tt 0xB3} \\ \hline
\end{longtable}

Pop a word, calculate the global port id, push it:

\begin{tt}
\noindent
a $\leftarrow$ m(sp++)\\
m(--sp) $\leftarrow$ port(thread, a)\\
ip $\leftarrow$ ip + 1
\end{tt}

\subsection{Calculate Stack Pointer}

\begin{longtable}{|p{2.5in}|p{2.5in}|}
\hline
Name: LDAX &
Opcode: {\tt 0xB4} \\ \hline
\end{longtable}

Pop an index, add it to the stack pointer, push the data pool index for
the word it points to:

\begin{tt}
\noindent
a $\leftarrow$ m(sp++)\\
a' $\leftarrow$ a + sp - dp\\
m(--sp) $\leftarrow$ a'\\
ip $\leftarrow$ ip + 1
\end{tt}

\subsection{Available Threads}

\begin{longtable}{|p{2.5in}|p{2.5in}|}
\hline
Name: THREADS &
Opcode: {\tt 0xB5} \\ \hline
\end{longtable}

Determine the number of threads available to start:

\begin{tt}
\noindent
m(--sp) $\leftarrow$ $\sum_{i=0..7}({\neg}started_i)$\\
ip $\leftarrow$ ip + 1
\end{tt}

\subsection{Cycles Per Threads}

\begin{longtable}{|p{2.5in}|p{2.5in}|}
\hline
Name: THRCYC &
Opcode: {\tt 0xB6} \\ \hline
\end{longtable}

Push the instruction cycle counter of the thread:

\begin{tt}
\noindent
m(--sp) $\leftarrow$ $cycles_t$\\
ip $\leftarrow$ ip + 1
\end{tt}

\subsection{Total Cycles}

\begin{longtable}{|p{2.5in}|p{2.5in}|}
\hline
Name: CYCLES &
Opcode: {\tt 0xB7} \\ \hline
\end{longtable}

Push the total instruction cycle counter of all threads of the
processing unit:

\begin{tt}
\noindent
m(--sp) $\leftarrow$ $\sum_{i=0..7}(cycles_i)$\\
ip $\leftarrow$ ip + 1
\end{tt}

\section{Implementation}

There is no hardware implementation of the
{\it Null Operand Parallel} processor,
but a software simulation.
As this simulator is to be taken as a proof of concept only,
optimisation efforts have been restricted to the bare minimum.
It may be given a number of command line parameters to take
influence on its detailed behaviour.
Next to a number of possible options, it is given a socket number
so it can handle a number of sockets, one for each of the four
external links. Furthermore, it may be given up to four
socket numbers of other simulator instances which it shall connect to:

~

\noindent
{\tt nopsim} [options] {\it first-own-socket} [{\it connect-to-socket} ...]

~

For a first compiler to support code generation, see \lref{2016os}.
However, there is no reason to not implement further compilers
for other languages.

Options to the simulator are:

\noindent
\begin{tabular}{|p{.3in}|p{1.2in}|p{3.4in}|} \hline
{\tt -i} & {\tt --init} {\it file} & initial code to read,
first five words skipped \\ \hline
{\tt -f} & {\tt --file} {\it file} & open file on peripheral link,
starting at third, i.e. at internal peripheral link \#2.
link \#0 is always connected to standard input,
link \#1 is always connected to standard output \\ \hline
{\tt -t} & {\tt --trace=}{\it mask} & full trace,
optionally restricted to threads: For the 32 bit {\it mask}, each single bit
represents a thread, starting at processing unit \#0, thread \#0,
processing unit \#0, thread \#1, and so forth, up to
processing unit \#3, thread \#7. For each bit set,
for the corresponding thread a full trace is sent to {\it stderr}.
The {\it mask} may be omitted to ask for trace of all threads \\ \hline
{\tt -x} & {\tt --extern=}{\it mask} & external links trace,
optionally for selected links.
Bit 0 to 3 represent the external links,
bit 4 to 11 represent peripheral lines,
bit 12 represents the router configuration block \\ \hline
{\tt -l} & {\tt --intern=}{\it mask} & internal links trace,
optionally restricted to threads.
The {\it mask} to be used as with {\tt --trace} \\ \hline
{\tt -d} & {\tt --debug} & enable break instruction and debug mode \\ \hline
{\tt -h} & {\tt --help} & show help and exit \\ \hline
\end{tabular}

~

Initially, the simulator will open sockets {\it first-own-socket} to
{\it first-own-socket}$+3$. For the first $n$ among these, with
$n$ the number of {\it connect-to-socket} parameters, it will attempt
to connect to the latter sockets. For the remaining $4-n$ sockets,
i.e. {\it first-own-socket}$+n$ to {\it first-own-socket}$+3$,
it will wait for another simulator to connect to it.
Only when all four sockets are connected, the simulator will start
executing opcodes.

As a special case, a stand-alone system may be simulated
by omitting all parameters.

Each of the processing units will run code from a simulated boot ROM
at address {\tt 0x3fc0}, which contains the following instructions
intended to read some initial code from its first internal port
(see option {\tt --init}):

~

\begin{tabular}{p{1.2in}p{3.7in}}
{\tt 0 IN} & read initial instruction word position \\
{\tt -64} & initial memory reference (first word) \\
{\tt 0 INMORE} & check for end of input \\
{\tt 10 FJP} & when done, skip to {\tt POP} below \\
{\tt DUP} & make use of memory reference \\
{\tt 0 IN} & read a word \\
{\tt EXCH ST} & store it to memory \\
{\tt 1 ADD} & increment memory reference \\
{\tt -12 UJP} & loop for next word \\
{\tt POP} & discard memory reference \\
{\tt 4 MUL JUMP} & transfer control to newly loaded code \\
\end{tabular}

\section{Discussion}

The main purpose of the {\it Null Operand Parallel} processor
is to allow for a prototypical implementation
of a dynamic software system as a proof of concept,
and its design is restricted to features
that are needed to serve this purpose.

For instruction encoding a single operation stack oriented approach
without registers -- occasionally referred to as {\it bytecode} --
has been chosen, to achieve compact encoding,
and to allow for a most simple code generation,
reducing effort in compiler implementation.
Clearly there are disadvantages, e.g. code execution is slower
than with a multiple register encoded instruction set, because
extra instructions are needed to fetch and store operands.
Other stack oriented designs have been proposed, that implement
multiple operations per instruction,
with the overall number of instructions reduced
as to allow for more compact implementation in hardware
(e.g. \lref{2010jb}).

Program flow exceptions have been replaced by exception messages,
but these need to be reduced to a minimum, as a process should
care for as much exceptional states as possible on its own.
Exceptional states may be divided into two classes, one class
are exceptional states that arise due to wrong encoding,
e.g. illegal instructions, resource inavailabilities,
stack overflow and other range check faults.
These exceptions need external handling, so it is inevitable
to send a message to some responsible process.
The second class are exceptional states that arise due to
unexpected input data,
e.g. division by zero, and sudden reception of an end token
where a message was not expected to end.
These exceptions should all be handled by the process itself.
For sudden reception of an end token, the instruction encoding
could be chosen to force evaluation of the token type
received by the process, e.g. by extending the input instruction
to return an extra boolean to indicate the type of the token
received.
For division by zero, a similar approach could be taken,
e.g. by extending the division instruction to return an extra
boolean to indicate successful operation, or by simply redefining
the division operation to result in a defined value whenever
the denominator is zero. The latter still requires the process
to do some extra check on the denominator first, but this is
not worse than the traditional exception based handling.

Directly connected to exception handling is the question
of how to respond to processes dying prematurely, resulting
in ports an thus channels to be no longer available.
For channels used for outgoing messages, any pending message
is finalised with an end token, but for channels used for
incoming messages, the process sending a message would need
to be notified about the fact that the destination has vanished.

Support for floating point calculation has not been implemented,
as it is not needed for basic operating system design.

True general purpose I/O lines are not foreseen, because an
implementation in hardware is not planned with this first design.
There are other designs that prove smooth integratability
into a channel based design is possible (e.g. \lref{2009dm}).

\newpage

\subsection*{Literature}

\def\Lit#1#2{\item {\tt [#1]}
#2}
\begin{list}{}{
  \setlength{\labelwidth}{0mm}
  \setlength{\itemsep}{0ex plus0.2ex}
  \setlength{\leftmargin}{6ex}
  \setlength{\itemindent}{-6ex}
  \setlength{\labelsep}{0mm}}

\Lit{1963bc}{
  Burroughs Corporation:
  ``The Operational Characteristics of the Processor for the Burroughs B5000'',
  1963,
  Detroit, Michigan
}

\Lit{2001dl}{
  Daan Le{\ij}en:
  ``Division and Modulus for Computer Scientists'',
  University of Utrecht,
  Dept. of Computer Science,
  December 3, 2001
}

\Lit{2007el}{
  Charles Eric LaForest:
  ``Second-Generation Stack Computer Architecture'',
  April 2007,
  University of Waterloo, Canada
}

\Lit{2008es}{
  ECSS Secretariat:
  ``SpaceWire -- Links, nodes, routers and networks'',
  ECSS-E-ST-50-12C, 31~July~2008, ESA-ESTEC.
}

\Lit{2009dm}{
  David May:
  ``The XMOS XS1 Architecture'',
  Version~1.0,
  XMOS Ltd., 2009/10/19
}

\Lit{2010jb}{
  James Bowman:
  ``J1: a small Forth CPU Core for FPGAs'',
  Willow Garage, Menlo Park, CA,
  2010
}

\Lit{2011ga}{
  GreenArrays, Inc.:
  ``F18A Technology Reference -- Product Data Book'',
  \scalebox{.76}[1]{http://www.greenarraychips.com/home/documents/greg/DB001-110412-F18A.pdf},
  12~April 2011
}

\Lit{2016os}{
  Oskar Schirmer:
  ``GuStL -- An Experimental Guarded States Language'',
  Göttingen, 2016
}

\end{list}

\end{document}